\documentclass[12pt,a4paper,twoside,english]{article}
     \usepackage[latin1]{inputenc}
     \usepackage[T1]{fontenc}
     \usepackage[usenames,dvipsnames]{color}
     \usepackage{graphicx}
     \usepackage{babel}
     \usepackage{html}
     \usepackage{verbatim}
     \usepackage{amsmath}
     \usepackage[nomarkers]{endfloat}
     \usepackage{epsf,rotate,a4wide}
\newcommand{\lp}{\left(}
\newcommand{\rp}{\right)}
\newcommand{\lc}{\left[}
\newcommand{\rc}{\right]}

\newcommand{\dr}{\partial}
\newcommand{\bfr}{{\bf r}}

\newcommand{\pca}{{\cal P}}

\addtocontents{lof}{\protect\vspace{0.5cm}}

\newcommand{\tba}{\overline{T}}

\newcommand{\pib}{\overline{\pi}}

\newcommand{\hba}{\overline{H}}
\newcommand{\pba}{\overline{p}}
\newcommand{\phib}{\overline{\phi}}
\newcommand{\gammaba}{\overline{\gamma}}
\newcommand{\gammast}{\gamma^*}

\newcommand{\lamba}{\overline{\lambda}}

\newcommand{\nd}{{\sf d}}

\newcommand{\clst}{{\cal L}^*}
\newcommand{\clba}{\overline{\cal L}}
\newcommand{\cxst}{{\cal X}^*}
\newcommand{\cxba}{\overline{\cal X}}
\newcommand{\cxha}{\widehat{\cal X}}
\newcommand{\cx}{{\cal X}}
\newcommand{\cz}{{\cal Z}}

\newcommand{\Real}{{\rm I \! \! \! \; R}}
\newcommand{\Complex}{{\rm \, \rule[0.1mm]{0.07mm}{2.5mm} \! \! \! \; C}}

\newcommand{\nit}{n}

\begin{document}

\newlength{\figwidth}
\setlength{\figwidth}{0.9\textwidth}

%

\vspace*{0.5cm}
\normalsize

\centerline{\bf Stability of Semi-Implicit and Iterative Centred-Implicit Time Discretisations}
\medskip
\centerline{\bf for Various Equation Systems Used in NWP}

\bigskip
\bigskip
\centerline{(Accepted in Monthly Weather Review)}
\bigskip

\rm

\centerline{\sc P. B{\'e}nard$^*$}
\bigskip
\bigskip
\centerline{\footnotesize $^*$ \sl Centre National de Recherches Météorologiques, Météo-France, Toulouse, France}
\medskip
\bigskip
\bigskip

\rm
\vspace{1in}
\centerline{27 March 2003}
\vspace{1in}
Corresponding address:

\medskip

Pierre Bénard

CNRM/GMAP

42, Avenue G. Coriolis

F-31057 TOULOUSE CEDEX

FRANCE

\bigskip

Telephone: +33 (0)5 61 07 84 63

Fax: +33 (0)5 61 07 84 53

e-mail: pierre.benard@meteo.fr

\newpage

\centerline {ABSTRACT}
\bigskip

The stability of classical semi-implicit scheme, and some more 
advanced iterative schemes recently proposed for NWP purpose
is examined.
In all these schemes, the solution of the centred-implicit 
non-linear equation is approached by an iterative fixed-point algorithm, 
preconditioned by a simple, constant in time, linear operator.
A general methodology for assessing analytically the stability 
of these schemes on canonical problems for a vertically 
unbounded atmosphere is presented. 
The proposed method is 
valid for all the equation systems usually employed in NWP.
However, as in earlier studies, the method can be applied 
only in  simplified meteorological contexts, thus overestimating 
the actual stability that would occur in more realistic meteorological 
contexts.
The analysis is performed in the spatially-continuous 
framework, hence allowing to eliminate the spatial-discretisation
or the boundary conditions as possible causes of the 
fundamental instabilities linked to the time-scheme itself.
The general method is then shown concretely to apply to various
time-discretisation schemes and equation-systems 
(namely shallow-water, and
fully compressible Euler equations).
Analytical results found in the literature are 
recovered from the proposed method, 
and some original results are presented.

\bigskip
\bigskip

\newpage

\section{Introduction}

The classical semi-implicit (SI) technique (Robert {\sl et al.}, 1972) 
has been widely used in NWP since it provides efficient and simple
algorithms, at least for spectral models. 
This classical SI method requires the definition of a constant
in time linear "reference" operator $\clst$, which usually consists
in the linearisation of the original system
${\cal M}$, around a stationary reference-state, noted $\cxst$.
For a given state ${\cal X}$ of the atmosphere, the evolution of the 
system, $(\dr {\cal X} / \dr t) = {\cal M}.{\cal X}$, 
is then time-discretised through:

\vspace{-0.4in} 
\begin{equation}
\frac{\delta {\cal X}}{\delta t} = ({\cal M} - \clst) .{\cal X} + \clst. \overline{[\cal X]}^t
\label{eq_MSI}
\end{equation} 
\vspace{-0.41in}


\noindent where $(\delta/ \delta t)$ is the discretised time-derivative
operator, and $\overline{[\rule[2mm]{2mm}{0mm}]}^t$ is the implicit-centred 
temporal average operator.
The terms linked to the reference operator $\clst$
are thus treated in a centred-implicit way, whilst the residual 
non-linear terms are treated explicitly.
For this scheme, there is no formal proof of the stability
in real atmospheric conditions, due to the explicit treatment 
of non-linear residuals. 
This prompted the authors of pioneering NWP 
applications of the SI scheme to examine theoretically its
stability in idealised contexts. 

In a seminal study following this approach, Simmons et al., 1978 
(SHB78 hereafter)
analysed the stability of the SI scheme for the hydrostatic primitive 
equations (HPE) system with a Leap-Frog (3-TL hereafter)
time-discretisation by considering  
the linearised equations around a stationary state 
$\cxba$ (referred to as "atmospheric state" hereafter) 
when the resulting linear "atmospheric" operator $\clba$
deviates from the linear "reference" operator $\clst$ of the SI scheme,
thus generating potentially unstable explicitly-treated residuals.

In the vertically-continuous context
they performed a stability analysis valid when the 
eigenfunctions of $\clba$ and $\clst$ are identical.
They showed that 
when the atmospheric and reference temperature 
profiles (respectively $\tba$ and $T^*$) are 
isothermal, the stability of the SI scheme requires:

\vspace{-0.4in} 
\begin{equation}
0 \leq  \tba \leq  2 T^*,
\label{eq_SHB}
\end{equation} 
\vspace{-0.41in}

\noindent hence $T^*$ cannot be chosen arbitrarily
for applying the 3-TL SI scheme to the HPE system.

In the finite-difference vertically-discretised context, 
thew showed that a "vertically-discretised analysis" of stability following
the same principle simply resulted in the solution of a standard 
eigenvalue problem.
They found empirically that a large static-stability for the 
reference-state is necessary to maintain the 
stability of the scheme for realistic thermal 
atmospheric profiles.
As a consequence, they recommended to use a warm isothermal
state as reference-state, a rule which was then widely adopted 
for NWP applications using SI schemes.

Finally, they examined the effect of applying a second-order 
time-filter in the temporal average of linear terms, and found that an 
improved stability is obtained, but at the expense of 
an increased misrepresentation of the wave propagation.

\medskip

Côté et al., 1983 (CBS83 hereafter), still in the 
HPE context, examined the stability of the
3-TL SI scheme for a finite-element vertical discretisation
using the above vertically-discretised analysis method.
They established a stability criterion for
the 3-TL SI scheme in terms of the atmospheric and reference
static stabilities ($\gammaba$ and $\gammast$ 
respectively):

\vspace{-0.4in} 
\begin{equation}
0 \leq \gammaba \leq  2 \gammast,
\label{eq_CBS}
\end{equation} 
\vspace{-0.41in}

\noindent therefore generalizing (\ref{eq_SHB}) to not 
necessarily isothermal thermal profiles. 

Still in the HPE context,
Simmons and Temperton, 1997  showed with the same method that
 extrapolating two-time level 
(2-TL) schemes have more stringent stability constraints 
than their 3-TL counterpart. For instance, in the isothermal framework of 
the SHB78 analysis, the stability of the 2-TL SI scheme requires:

\vspace{-0.4in} 
\begin{equation}
0 \leq  \tba \leq   T^*.
\label{eq_ST}
\end{equation} 
\vspace{-0.41in}

\noindent As a consequence, they recommended to use a warmer reference 
temperature than in the 3-TL case. 
ST97 also showed that the 2-TL SI scheme was intrinsically damping
when stable, a characteristic which was not present in the 3-TL SI
scheme. However, 3-TL schemes require a time-filter to damp the 
unstable computational modes, and this time-filter also
damps the transient physical modes. They argue that 
all these aspects being considered, the effective damping 
of 2-TL and 3-TL schemes is of comparable overall intensity.
This particular debate will be ignored in this paper, in order
to focus on stability aspects only.

The relevance of the SI method for solving numerically the fully 
compressible Euler Equations (EE) was then advocated (Tanguay et al., 1990), 
and some numerical models in EE using this technique were effectively
developed: Caya and Laprise (1999), presented a model in EE with a 3-TL SI scheme
(with a moderate time-decentering first-order accurate in time). 
Semazzi et al. (1995) and Qian et al. (1998) also showed a model in EE but with a 2-TL SI scheme
(with a strong time-filter however). 
Besides, the need of more robust schemes than the classical
SI one for solving the EE system became recognized (e.g. C{\^o}t{\'e}
et al. 1998, BHBG95), probably motivated by some
pathological behaviours with the classical SI scheme
under some circumstances.

Schemes with evolution terms treated in a more centred-implicit way are
usually believed to have an increased robustness, hence 
fulfilling the latter emerging need, and some 
of them were developed for fine-scale models in the EE system.
Bubnov\'a et al., 1995 used a 3-TL scheme in which the 
leading non-linear terms of the EE system are treated in a 
centred-implicit way, through a partially iterative  method.
For a 2-TL scheme, C\^ot\'e et al, 1998, used a fully iterative method, 
aiming to treat all the evolution terms of the EE system in a 
centred-implicit way.
Cullen (2000) examined the benefit of using such a fully iterative 
scheme for the HPE system, arguing that an improved accuracy could be obtained
beside the improved stability (this latter being not strictly
required however for current HPE applications).
As a formal justification, he examined the stability of this 
iterative scheme for the 2-TL shallow-water (SW) system. 
The analysis was limited to a scheme called "predictor/corrector",
which consists in a single additional iteration after the SI scheme.
The salient result was that the additional iteration in the
"predictor/corrector" scheme allows to recover a
extended range of stability as in (\ref{eq_SHB}) instead of (\ref{eq_ST}).
In the following, these fully iterative schemes with a more 
centred-implicit treatment of the evolution terms will be 
referred to as "iterative centred-implicit" (ICI) schemes.

From the theoretical point of view, the current situation 
is that no stability analysis has been
provided for the EE system with SI scheme, and for 
ICI schemes with more iterations, stability analyses 
are available only for the SW system.

Here we present a general method to carry out space-continuous
stability analyses of the various time-discretisation schemes 
mentioned above, for any usual meteorological system 
of NWP interest (SW, HPE, EE), on canonical problems
similar to those examined in SHB78.
Some original results concerning the EE system and iterative schemes 
are presented.

This work may also be viewed as a first theoretical 
investigation about the suitability of various 
time-discretisation schemes for solving numerically 
the EE system.


\section{General framework for analyses}
\label{sec_fram}

The general framework for the stability analyses presented 
here is basically the same as in most earlier studies: 
The flow is assumed adiabatic inviscid and frictionless
in a non-rotating dry perfect-gas atmosphere 
with a Cartesian coordinate system.
Moreover, the flow is assumed linear around an
"atmospheric" basic-state $\cxba$. The actual evolution of the 
atmospheric flow is thus described by $\clba$, 
the linear-tangent operator to ${\cal M}$ 
around $\cxba$. The atmospheric basic-state $\cxba$
is chosen stationary, resting, horizontally homogeneous, 
and hydrostatically balanced.  
The governing equation for the flow is then:

\vspace{-0.4in} 
\begin{equation}
\frac{\dr \cx' }{\dr t} = \clba . \cx'
\label{eq_lin_base}
\end{equation} 
\vspace{-0.41in}

\noindent where $\cx' = \cx - \cxba$, and the primes are dropped 
henceforth for clarity.

Following the usual practice in NWP, the linear operator 
$\clst$ in (\ref{eq_MSI}) is taken as the tangent-linear 
operator to ${\cal M}$ around a reference-state $\cxst$ which
is also chosen stationary, resting, horizontally homogeneous, 
and hydrostatically balanced. 
Since $\cxba$ is a resting state, the 
linear Lagrangian time-derivative coincides with the 
Eulerian time-derivative, and the LHS operator of  
(\ref{eq_lin_base}) hence holds for Eulerian models 
as well as for semi-Lagrangian models .

\section{The class of ICI schemes in the linear framework}
\label{sec_Class}

In the restricted resting and linear framework of section \ref{sec_fram},
classical SI schemes as well as the iterative schemes mentioned in the
introduction can be gathered in a single class of ICI schemes, differing
only by their number of iterations.
These ICI scheme are first presented for a 2-TL discretisation.
In this case, the fully implicit-centred (FIC) scheme writes:

\vspace{-0.4in} 
\begin{equation}
\frac{{\cal X}^{+} - {\cal X}^{0}}{\Delta t} 
 =  \frac{\clba.{\cal X}^{+} + \clba.{\cal X}^{0}}{2}
\label{eq_FIC2TL}
\end{equation} 
\vspace{-0.41in}

\noindent where, according to a standard practice for time-discretised 
equations, the superscripts "+" and "0"  
stand for time levels $(t+\Delta t)$ and $t$ respectively.
The principle of ICI schemes is to approach the FIC solution by starting
from an initial "guess" noted ${\cal X}^{+(0)}$, then iterating the 
following algorithm:

\vspace{-0.4in} 
\begin{eqnarray}
\frac{{\cal X}^{+(\nit)} - {\cal X}^{0}}{\Delta t} 
& = & \frac{\clba.{\cal X}^{+(\nit -1)} + \clba.{\cal X}^{0}}{2}
+ \frac{\clst.{\cal X}^{+(\nit)} - \clst.{\cal X}^{+(\nit-1)}}{2}
\label{eq_ICI2TL} \\
& \equiv & \frac{ \lp \clba - \clst \rp .{\cal X}^{+(\nit -1)} 
                + \lp \clba - \clst \rp .{\cal X}^{0}}{2}
+ \frac{{\clst}.{\cal X}^{+(\nit)} + {\clst}.{\cal X}^0}{2}
\label{eq_ICI2TLbis}
\end{eqnarray} 
\vspace{-0.41in}

\noindent for $\nit =1,2,...,N_{\rm iter}$. The $\cx^+$ state, 
valid a $(t + \Delta t)$ is then taken as the last iterated state
${\cx}^{+(N_{\rm iter})}$. 
 An examination of this scheme for a
model with a single prognostic variable without spatial dependency
shows that it acts as a fixed-point algorithm for solving the implicit
non-linear scalar equation $f(x)=x$, by using an estimate
${f'}^{*}$ of the derivative $f'$ as a preconditioner for convergence.
The method converges if $\left | (f' - {f'}^{*}) / {f'}^{*} \right | < 1$.
This is a weaker condition than the one for the classical 
(i.e. not preconditioned) fixed-point method: $\left | f' \right | < 1$.
The initial guess ${\cal X}^{+(0)}$ is arbitrary in ICI 
schemes, but choosing an appropriate initial guess may
help decreasing the magnitude of the discrepancy between 
FIC and ICI schemes after a fixed number of iterations.

\noindent For 3-TL schemes, the ICI scheme can be defined by:

\vspace{-0.4in} 
\begin{eqnarray}
\frac{{\cal X}^{+(\nit )} - {\cal X}^{-}}{2 \Delta t} 
& = & \frac{{\clba}.{\cal X}^{+(\nit -1)} + {\clba}.{\cal X}^{-}}{2}
+ \frac{{\clst}.{\cal X}^{+(\nit )} - {\clst}.{\cal X}^{+(\nit -1)}}{2}
\label{eq_ICI3TL}\\
& \equiv & \frac{ \lp \clba - \clst \rp .{\cal X}^{+(\nit -1)} 
                + \lp \clba - \clst \rp .{\cal X}^{-}}{2}
+ \frac{{\clst}.{\cal X}^{+(\nit)} + {\clst}.{\cal X}^-}{2}
\label{eq_ICI3TLbis}
\end{eqnarray} 
\vspace{-0.41in}

\noindent where the superscript "-" denotes a variable taken 
at the time-level $(t-\Delta t)$.

Here follows a list of some schemes proposed in the literature, with
the corresponding characteristics (${\cx}^{+(0)}$ ,$N_{\rm iter}$),
in the restricted framework of section \ref{sec_fram}:

\begin{list}{}{}
\item[-] Classical 2-TL SI extrapolating scheme: $N_{\rm iter}=1$ and
 ${\cx}^{+(0)} = (2 \cx^0 - \cx^-)$.
\item[-] 2-TL non-extrapolating SI scheme: $N_{\rm iter}=1$ and 
      ${\cx}^{+(0)} =  \cx^0$. However, this scheme is not used 
      in practice since it is only first-order accurate in time,
      as mentioned in Cullen (2000).     
\item[-] "Predictor/corrector" scheme of Cullen (2000): $N_{\rm iter}=2$ 
      and  ${\cx}^{+(0)} =  \cx^0$. 
\item[-] Iterative scheme of Côté et al., 1998: general iterative ICI scheme, 
      but used with $N_{\rm iter}=2$ in practice. The choice of
      ${\cx}^{+(0)}$ is not explicitly indicated.
\item[-] FIC scheme:  $N_{\rm iter}=\infty$ (does not depend 
      on the choice of ${\cx}^{+(0)}$). 
      This scheme can not be achieved in practice for numerical 
      models, but it may be useful for theoretical examination
      of the asymptotic behaviour of the ICI schemes.
\end{list}

\noindent For 3-TL discretisations, the SI scheme, which corresponds to an ICI scheme 
 with $N_{\rm iter}=1$ and ${\cx}^{+(0)} = (2 \cx^0 - \cx^-)$ 
 is the only one to be used in practice. However, iterated 3-TL ICI schemes 
 could be used as well, and the 3-TL FIC scheme is equivalent 
 to a 2-TL FIC scheme with a time-step twice as long.

In the general framework, when ${\cal M}$ is not 
linear, these various schemes cannot be gathered 
in the unique formalism 
(\ref{eq_ICI2TL}) or (\ref{eq_ICI3TL}).

Addition of a second-order time-filter 
in the above definitions for 2-TL schemes is straightforward.
Two main variants are usually considered, 
depending on whether the filtering is applied
only to the time-averages of linear terms (as in SHB78),
or also to the time-averages of non-linear terms,
in (\ref{eq_ICI2TLbis}).   
For instance, in a 2-TL SI scheme, the first variant 
of the filter consists in replacing $\clst(\cx^{+(n)} + \cx^0)$
by $\clst[(1 + \kappa) \cx^{+(n)} + (1 - 2 \kappa) \cx^0 + \kappa \cx^-]$ 
in (\ref{eq_ICI2TLbis}), where $\kappa$ is a (small) positive parameter, 
and for the second variant, the same modification is also applied to the
first RHS term of (\ref{eq_ICI2TLbis}).
The scheme then becomes essentially a 3-TL scheme since information at level
$\cx^-$ is always used. 
However, the use of large values of $\kappa$ (e.g. $\kappa=0.5$,
which eliminates the $\cx^0$ contribution) 
is known to deteriorate the solution through a spurious damping
of transient perturbations (e.g. Hereil and Laprise, 1996).

For 3-TL schemes, second-order time-filters are uneffective, and
a first-order accurate time-decentering must be used.
This consists in replacing $\clst(\cx^{+(n)} + \cx^-)$ 
by $\clst[(1 + \epsilon) \cx^{+(n)} + (1 - \epsilon) \cx^-]$
in (\ref{eq_ICI3TLbis}), where $\epsilon$ is a (small) positive parameter.
The scheme then ceases to be second-order accurate in time
since the time-average is no longer centred in time.
This results in a spurious damping of transient perturbations
even for moderate values of $\epsilon$ (due to the weaker
time-selectivity of the filter $\epsilon$ compared
to $\kappa$).

\section{Conditions for space-continuous analyses}
\label{sec_cond}

\subsection{Conditions on the upper and lower boundaries}

Space-continuous analyses are much easier to carry out 
when the equation system is defined in the whole 
unbounded atmosphere, because the expression of the 
normal modes of the system is more general.
The following space-continuous analyses will
restrict to this case (although this is not strictly 
required).
For systems in which the vertical direction is represented 
(e.g. HPE and EE systems),
this means that the upper and lower boundary conditions
must not appear explicitly in the set of governing equations. 
However for systems cast in mass-based coordinates 
[such as HPE system in pressure-based coordinate (e.g. SHB78), and
EE system in hydrostatic pressure-based coordinate 
(Laprise, 1992, L92 hereafter)],
the upper and lower boundary conditions actually appear inside the 
set of equations through vertical integral operators
with definite bounds at the boundaries of the vertical domain.
When they are present, it is assumed that these integral 
operators can be eliminated, i.e.
that $\clba$, $\clst$  can be transformed to 
"unbounded" operators by application of appropriate 
vertical linear differential operators to the prognostic equations 
which originally involve integral operators, in order that e.g.
(\ref{eq_lin_base}) rewrites as:

\vspace{-0.2in} 
\begin{equation}
\frac{\dr}{\dr t}
\lp
\begin{array}{c}
l_{1} \cx_1 \\ \vdots \\ l_{P} \cx_P
\end{array}  \rp = \lp
\begin{array}{ccc}
l_1 \clba_{11}  & \cdots & l_P \clba_{1P}  \\
\vdots & \ddots & \vdots \\
l_1 \clba_{P1}  & \cdots & l_P \clba_{PP}  
\end{array} 
\rp .  \lp 
\begin{array}{c}
\cx_1 \\ \vdots \\ \cx_P
\end{array} 
\rp
\label{eq_linder_base_var}
\end{equation} 
\vspace{-0.11in}

\noindent where $P$ is the number of prognostic variables of the unbounded system,  
$(l_{1},\ldots,l_{P})$ are linear vertical operators, and
$(l_1 \clba_{11},\ldots, l_P \clba_{PP})$ are linear spatial operators which 
no longer contain any reference to the upper and lower boundaries.
The transformed system obtained for $\clba$
can then be written as:

\vspace{-0.4in} 
\begin{equation}
\frac{\dr l. \cx}{\dr t}  =  l. \clba . \cx   \label{eq_linder_base}
\end{equation} 
\vspace{-0.41in}

\noindent where $l$ is the diagonal matrix $(l_{1},\ldots,l_{P})$.
A similar condition must be true for  $\clst$ as well:
it is assumed that applying the same  
operator $l$ to $\clst$ leads to an operator $l \clst$ for which
$l_i \clst_{ij}$ does not contain any reference to the 
upper and lower boundaries for $(i, \; j) \; \in (1, \ldots, P)$. 
The first condition for the following analyses is:

\begin{list}{}{}
\item[] [C1]: There exists a linear operator $l$ such as $l \clba$ and $l \clst$ have no
reference to the upper and lower boundaries.
\end{list}

\noindent The system (\ref{eq_linder_base}) is henceforth referred to 
as the "unbounded" system.

\subsection{Conditions on the stability of the $\cxba$ state}

The aim of these analyses is to determine in
which conditions a stationary state $\cxba$ for $l \clba$ will remain 
a stable equilibrium-state in the time-discretised context, provided it is a 
stable equilibrium-state in the time-continuous context. 
Hence the analyses will be restricted to stationary states 
$\cxba$  which are in stable equilibrium.
Given the linear context used here, a physical transposition 
of this condition is:

\begin{list}{}{}
\item[] [C2]: For any perturbation $\cx(t=0)$ around $\cxba$ with a 
bounded energy-density, the time-evolution $\cx(t)$ resulting from
(\ref{eq_linder_base}) must have a bounded energy-density.
\end{list}

\noindent The condition is formulated with energy-density instead of
total energy because the domain is unbounded in space. 
The complex eigenmodes of the unbounded
system (\ref{eq_linder_base}) are the complex functions of space 
$\cx(\bfr)$ which satisfy:

\vspace{-0.4in} 
\begin{equation}
l \clba \cx(\bfr) = \lamba l \cx(\bfr)
\label{eq_eigen}
\end{equation} 
\vspace{-0.41in}

\noindent where $\lamba \in \Complex$ and 
$\bfr$ denotes the spatial dependency. The time-evolution of the mode 
is bounded in time if $\lamba$ is a pure imaginary number, 
and $\cx(\bfr)$ is then a "normal mode" 
in the usual terminology.
Theoretical arguments out of the scope of this paper show 
that a mathematical transposition of [C2] writes:

\begin{list}{}{}
\item[] [C2']: For any complex eigenmode of $l \clba $,
$\;\; \lamba \in i \Real \; \Longleftrightarrow \; \cx(\bfr)$ has a bounded energy-density.
\end{list}

\subsection{Conditions on the normal modes of the linear unbounded system}

\noindent The time-continuous normal modes of the original 
system (\ref{eq_lin_base}) are the complex functions of space 
$\cx(\bfr)$ which have an oscillatory temporal 
evolution. Hence they satisfy $\clba \cx(\bfr) = i \, \overline{\omega} \cx(\bfr)$.
Similarly, the time-continuous normal modes of the linear unbounded 
system (\ref{eq_linder_base}) are the complex functions of space 
$\cx(\bfr)$ for which $l \cx(\bfr)$ has an oscillatory temporal 
evolution. Hence they satisfy
$ l. \clba \cx(\bfr) = i \, \overline{\omega} l.\cx(\bfr)$.

\noindent Any normal mode of the original system is also 
necessarily a normal mode of the unbounded system, with
the same frequency $\overline{\omega}$.
For any normal mode of the unbounded system , we can choose 
the origin of space {\bf o} in such a way that this mode writes:

\vspace{-0.4in} 
\begin{equation}
\cx(\bfr) =  \cx({\bf o}) f(\bfr) = \cxha f(\bfr)
\label{eq_mode}
\end{equation} 
\vspace{-0.41in}

\noindent with
$\cxha = (\cxha_1, \ldots, \cxha_P) \in  \Complex^P$, and
$f=(f_1, \ldots, f_P)$ is a vector of space-dependent functions
(the product $\cxha f \;$ is understood "component by component").
For the indices $i$ such as $\cx_i$ uniformly vanishes, 
$\cxha_i=0$ and $f_i(\bfr)=1$ are assumed by convention.
The function $f$ and the vector $\cxha$ are respectively 
termed the "structure" and the "polarisation vector" of the mode.
The two following conditions are required for the proposed analyses:

\begin{list}{}{}
\item[] [C3]: For any normal mode $\cx$ of the unbounded linear 
atmospheric system with a structure $f(\bfr)$, $\; l_{i} . f_i(\bfr)$ 
must be proportional to $f_i(\bfr)$:
\end{list}

\vspace{-0.8cm}

\vspace{-0.4in} 
\begin{equation}
\forall i \in (1, \ldots ,P) {\rm \; , \;\;} l_i f_i(\bfr) = \xi_i f_i(\bfr)
{\rm \;\; with \;} \xi_i \in \Complex^*.
\label{eq_cnt}
\end{equation} 
\vspace{-0.41in}

\noindent and:

\begin{list}{}{}
\item[] [C4]: For any normal mode $\cx$ of the unbounded linear 
atmospheric system with a structure $f(\bfr)$,
$\; \clba_{ij} . f_j(\bfr)$ 
[resp. $\clst_{ij} . f_j(\bfr)$]  must be 
proportional to $f_i(\bfr)$:
\end{list}

\vspace{-0.8cm}

\vspace{-0.4in} 
\begin{equation}
\forall (i,j){\rm ,\;\; }  
l_i \clba_{ij} . f_j(\bfr) = \overline{\mu}_{ij} f_i(\bfr) \;\;  {\rm and \;\; } 
l_i \clst_{ij} . f_j(\bfr) = \mu^*_{ij} f_i(\bfr) 
{\rm , \;  with \;\;} (\overline{\mu}_{ij}, \, \mu^*_{ij}) \;\;{\in} \;\; \Complex.
\label{eq_cnq}
\end{equation} 
\vspace{-0.41in}

\noindent As will be seen below, these latter two conditions have the important 
consequence that for any normal mode of the unbounded system,
each individual time-discretised prognostic equation for $\cx_i(\bfr)$
becomes a scalar equation. This key ingredient makes the analysis straightforward
for every member of the ICI class.

\subsection{Comments}

Since the set of normal modes for the unbounded system encompasses the set of 
normal mode of the bounded system, the transformation from the bounded system to 
the unbounded system is not likely to "mask" some instabilities of
the original system, unless the causes
of the instability lie in the boundary conditions themselves. 
However, discretised-analyses allow to clarify this point
by showing that the stability of the bounded and unbounded 
systems are actually found to be similar in practice.
Besides, discretised-analyses of the unbounded system
are by nature impossible to perform, hence the continuous 
analysis is the only way to
estimate the intrinsic stability of the unbounded system and
to demonstrate that instabilities, when they occur in a practical
application, are not due to a weakness in the spatial discretisation
or even to the boundary conditions, but actually to the
time-discretised propagation of free modes inside the atmosphere.

In spite of their apparently abstract and constraining form, 
conditions [C1]--[C4] are easy to verify with routine
normal mode analysis techniques when examining 
a particular concrete meteorological system. 
The condition [C2'] restricts the set of stationary states
$\cxba$ around which the analysis is meaningful, and conditions [C1], [C3], [C4]
restrict the spectrum 
of meteorological contexts accessible to the analysis,
since they require a qualitative similarity between 
$l$, $\clba$ and $\clst$ operators. 
As stated in SHB78, analyses performed under 
this type of conditions "...grossly exaggerate the 
stability of the scheme..." since in more realistic 
meteorological contexts, the atmospheric and reference 
operators can be qualitatively much more different 
than imposed  by this condition.

\section{Time-Discretised Space-Continuous Analysis}
\label{sec_tdana}

The analysis examine the stability of the
time-discretised system for perturbations which have a time-continuous normal 
mode structure.
Hence we consider a given function $f=(f_1, \ldots, f_P)$ which 
is a normal mode structure for the time-continuous system, 
and we determine the normal modes of the time-discretised
system which have the same structure $f$, by solving the 
equation:

\vspace{-0.4in} 
\begin{equation}
\cxha_{(t=\Delta t)} f({\bf r})  = \lambda \cxha_{(t=0)} f({\bf r})
\label{def_lambda}
\end{equation} 
\vspace{-0.41in}

\noindent where $\cxha_{(t=0)}$ and $\lambda$ are the unknowns, and
$\cxha_{(t=\Delta t)}$ is  determined using the time-discretisation scheme
(\ref{eq_ICI2TL}) or (\ref{eq_ICI3TL}). 
For schemes using three time levels (as Leap-Frog or extrapolating
2-TL) a similar relationship 
$\cxha_{(t=-\Delta t)} f({\bf r})  = \lambda^{-1} \cxha_{(t=0)} f({\bf r})$
must be added.
If for some solution, $\left | \lambda \right | > 1$ (resp. $<1$)
the scheme is unstable (resp. damping) for this particular mode. 
The ratio ${\rm Arg} ( \lambda)/ (i \overline{\omega} \Delta t)$
gives the relative  phase-speed error of the scheme 
for this mode.

The analysis is described here for a 2-TL discretisation
(\ref{eq_ICI2TL}), but the transformation to a 3-TL 
scheme as well as the addition of 
time-filters such as $\kappa$ or $\epsilon$ are straightforward. 
In the following of this section, the notation $\cx(\bfr,t)$ is 
replaced by the usual superscript notation for time-discretised 
variables $\cx^t(\bfr)$ as in section \ref{sec_Class}.

As a consequence of the discussion in section \ref{sec_Class} 
and applying (\ref{def_lambda}) ,
$\cx^{+(0)}$ can be written as
$ \cx^{+(0)}(\bfr) = \mu(\lambda) \cx^0(\bfr)$, 
where $\mu(\lambda)$ depends on the choice of the initial
guess $\cx^{+(0)}$ (e.g. $\mu =1$ for a 2-TL non-extrapolating scheme, 
$\mu =2 - 1/ \lambda$ for a 2-TL extrapolating scheme,  etc...).
The original unbounded system (\ref{eq_linder_base}) 
is thus time-discretised following (\ref{eq_ICI2TL}):

\vspace{-0.4in} 
\begin{eqnarray*}
l \cx^{+(0)}(\bfr) & = &  \mu(\lambda) \, l \cx^0(\bfr) \\
\frac{{l \cx}^{+(\nit )}(\bfr) - {l \cx}^{0}(\bfr)}{\Delta t} 
& = & \frac{{l \clba}.{\cx}^{+(\nit -1)}(\bfr) + {l \clba}.{\cx}^{0}(\bfr)}{2} \\
&  & + \frac{{l \clst}.{\cx}^{+(\nit )}(\bfr) - {l \clst}.{\cx}^{+(\nit -1)}(\bfr)}{2} 
\; , \;\; \nit \in (1, \ldots, N_{\rm iter})\\
l \cx^+(\bfr) \equiv l \lambda \cx^0 & = & l \cx^{+(N_{\rm iter})}(\bfr)
\end{eqnarray*} 
\vspace{-0.41in}

\noindent For schemes with $N_{\rm iter} \geq 2$, it is assumed 
that the intermediate states $l {\cx}^{+(\nit )}(\bfr)$ for 
$i \; \in (1,\ldots,N_{\rm iter}-1)$ have 
the same structure $f$ as the one currently examined,
which allows to define the polarisation vector $\cxha_i^{+(\nit)}$ by:

\vspace{-0.4in} 
\begin{equation}
l_i \cx_i^{+(\nit)} (\bfr) = \cxha_i^{+(\nit)} l_i f_i(\bfr).
\end{equation} 
\vspace{-0.41in}

\noindent Hence, using (\ref{eq_cnt}) and (\ref{eq_cnq}),
the space dependency $f_i(\bfr)$ eliminates. For the generalized state-vector 
$\cz=(\cxha^0,\cxha^{+(0)},\cxha^{+(1)},\ldots, \cxha^{+(N_{\rm iter})} )$
of length $(N_{\rm iter}+2 ) \times P$, the above system writes:

\vspace{-0.2in} 
\begin{equation}
\lp
\begin{array}{cccccc}
\mu(\lambda) I_P  & -I_P   &   0_P   & \cdots  & \cdots  & 0_P    \\
      M_1         &   M_2  &   M_3   & \ddots  &         & \vdots \\
    \vdots        &   0_P  &  \ddots & \ddots  & \ddots  & \vdots \\
    \vdots        & \vdots &  \ddots & \ddots  & \ddots  & 0_P    \\
        M_1       & \vdots &         & \ddots  &   M_2   & M_3    \\
- \lambda I_P     &   0_P  & \cdots  & \cdots  &   0_P   & I_P 
\end{array}
\rp . \cz  = {\bf M}. \cz = 0.
\label{eq_defm}
\end{equation} 
\vspace{-0.21in}

\noindent where $I_P$ (resp. $0_P$) is the unit (resp. null) $P$-order matrix, and:

\vspace{-0.4in} 
\begin{eqnarray}
\lp M_1 \rp_{ij} & = & - \delta_{ij} -   \frac{\Delta t}{2} \frac{\overline{\mu}_{ij}}{\xi_i} \\
\lp M_2 \rp_{ij} & = & - \frac{\Delta t}{2} \frac{1}{\xi_i} \lp \overline{\mu}_{ij} - \mu^*_{ij} \rp \\
\lp M_3 \rp_{ij} & = & + \delta_{ij}  -   \frac{\Delta t}{2} \frac{\mu^*_{ij}}{\xi_i} 
\end{eqnarray} 
\vspace{-0.41in}

\noindent where $\delta_{ij}$ is the $(i,j)$ Kronecker symbol.
The possible values of $\lambda$ for the normal mode structure
$f(\bfr)$ that we examine, are thus given by the roots of
the following polynomial equation in $\lambda$:

\vspace{-0.4in} 
\begin{equation}
{\rm Det} ( {\bf M}) = 0
\label{eq_detm}
\end{equation} 
\vspace{-0.41in}

\noindent  The dependencies to $\lambda$ are limited to the top- 
and bottom-left blocks.
For a non-extrapolating 2-TL scheme the degree of the polynomial
is $P$, and there are $P$ physical modes associated with this structure
$f(\bfr)$. For time-schemes making use of three time-levels (i.e.
3-TL schemes, or extrapolating 2-TL schemes, or 2-TL schemes with 
a time-filter), 
the degree becomes $2P$, and there are $P$ additional computational modes.
The growth-rate for any of these modes is given 
by the modulus of the corresponding complex root of (\ref{eq_detm}).
The growth rate of the time scheme for the considered structure $f$
is then defined by the maximum value of the modulus of these $P$ or $2P$ roots:

\begin{equation}
\Gamma(f) = {\rm Max} \lp \left |  \lambda_i(f) \right |\rp, 
\;\; i \in (1,\ldots, P) \; \; \lc \, {\rm or} \; \; (1, \ldots, 2P) \, \rc
\end{equation}

\noindent Analytical solution of (\ref{eq_detm}) is not possible for large values 
of $P$, and a numerical solution is often needed.

In this paper we will call "asymptotic growth-rate" the growth-rate
for the matrix ${\bf M}$ in the limit of large time-steps 
$\Delta t \rightarrow \infty$. 
The analysis of the asymptotic growth-rate is easier
than for finite time-steps,  since the matrix ${\bf M}$
of (\ref{eq_detm}) degenerates to a matrix ${\bf M'}$ 
in which $\delta_{ij}$ vanishes and $\Delta t/2$ eliminates.
Another advantage of asymptotic growth-rates is that they appear
to be independent of the structure $f$ in most of the cases 
examined below, thus simplifying considerably the interpretation
of the results.
When the asymptotic growth-rate is independent of the structure $f$
the growth-rate of the scheme can be defined by the growth-rate
obtained for this scheme with any structure $f$.

When the growth rate for a given scheme is one (or less) for any  
mode of any normal mode structure $f$, the scheme is then said to
be "unconditionally stable" (being understood "in $\Delta t$").
The criterion for unconditional stability obtained through 
${\rm Det} ( {\bf M'}) = 0$ is not 
only of academic interest since the considered time-schemes 
are actually used with large time-steps in NWP: 
the practice shows that
a scheme which is not unconditionally stable in the simplified context 
of these analyses has few chance to be robust enough for  
use in real conditions.

\section{Simple examples: 1D systems}
\label{sec_simex}

\subsection{1D Shallow-water system}

The linearised 1D shallow water system in an horizontal direction $x$
can be classically written in terms of the wind $u$ along $x$, and 
the geopotential $\phi$:

\vspace{-0.4in} 
\begin{eqnarray}
\frac{\dr u}{\dr t} & = &  - \frac{\dr \phi}{\dr x} 
\label{eq_exba}\\
\frac{\dr \phi}{\dr t} & = & - \phib \frac{\dr u}{\dr x} 
\label{eq_exbb}
\end{eqnarray} 
\vspace{-0.41in}

\noindent This system is also valid for the external mode of 
an isothermal atmosphere in HPE and EE system, replacing 
$\phib$ by $4 (R^2/C_p) \tba$ (which is done in the following).
The reference system is obtained by replacement of $\tba$ by $T^*$, and 
a "non-linearity" factor is defined through: $\alpha = (\tba - T^*)/T^*$.
Solution of (\ref{eq_eigen}) implies that if
$\tba < 0$, $\; \lamba \in i \Real  \; \Longrightarrow  \; u = \widehat{u} \exp (r x)$, 
which has not a bounded energy-density for $ r x \longrightarrow + \infty$.
Hence [C2'] requires $\tba \geq  0$ (i.e. $\alpha \geq  -1$).
The boundary conditions do not appear explicitly in the system, 
hence $l$ can be taken as the identity operator to satisfy [C1], and [C3].
In the notations of section \ref{sec_cond} and \ref{sec_tdana} 
we have $P=2$ , $\cx_1 = u$ and $\cx_2 = \phi$. 
The  normal modes of the system write $\psi(x)  =  \widehat{\psi} \exp(i k x)$
with $k \in \Real$ and $\psi=(u, \phi)$. Conditions [C1] -- [C4] are 
easily checked to be satisfied.

For a 3-TL SI scheme, (\ref{eq_detm}) writes:

\vspace{-0.4in} 
\begin{equation}
\lp \frac{\lambda^2 -1}{2 \Delta t} \rp^2 
= - \frac{k^2 c^{*2}}{4} 
\lp \lambda^2 + 1 \rp 
\lp \lambda^2 + 1 + 2 \alpha \lambda \rp,
\label{eq_SW_TTL}
\end{equation} 
\vspace{-0.41in}

\noindent where $c^*= 2 \sqrt{(R/C_p) R T^*}$. 
In the limit of long time-steps, the LHS term disappears, and
the four roots of the RHS give the "asymptotic" numerical growth-rate for the
two physical and two computational modes of the system.
The two roots of the first factor have a neutral stability, 
while those of the second factor have a modulus equal to 1
if $-1\leq  \alpha \leq 1$.
The criterion (on $\cxba$, $\cxst$) for unconditional stability
(in $\Delta t$) of the 3-TL SI scheme is thus : $0 \leq  \tba \leq  2 T^*$.
Some further algebraic manipulations from (\ref{eq_detm}) with 
$\Delta t = \infty$  show that this criterion remains unchanged
when increasing $N_{\rm iter}$.

For a 2-TL SI non-extrapolating ($\mu=1$) scheme, (\ref{eq_detm}) 
becomes:

\vspace{-0.4in} 
\begin{equation}
\lp \frac{\lambda -1}{\Delta t} \rp^2 
= - \frac{k^2 c^{*2}}{4} 
\lp \lambda + 1 \rp 
\lp \lambda  + 1 + 2 \alpha   \rp,
\label{eq_SW_DTL}
\end{equation} 
\vspace{-0.41in}

\noindent and the criterion for unconditional stability 
becomes more constraining than for the 3-TL scheme: $-1 \leq  \alpha \leq  0$ (i.e.
$0 \leq  \tba \leq  T^*$).
The asymptotic growth-rate of a 2-TL non-extrapolating
ICI scheme with $N_{\rm iter}$ iterations is given by:

\vspace{-0.4in} 
\begin{equation}
\Gamma = {\rm Max} \lp 1, \left |  2 (- \alpha)^{N_{\rm iter}} -1 \right | \rp
\label{eq_Lam}
\end{equation} 
\vspace{-0.41in}

\noindent The domain for unconditional stability is thus $-1 \leq  \alpha \leq  0$
for odd values of $N_{\rm iter}$, and $-1 \leq  \alpha \leq 1$ for even
values of $N_{\rm iter}$.

\subsection{1D vertical acoustic system in mass-based coordinate}
\label{sec_acoumass}

We consider a vertical 1D compressible atmospheric column satisfying
the conditions listed in section \ref{sec_fram}.
A regular mass-based coordinate $\sigma = (\pi / \pi_s)$  is chosen
following L92, by making $\eta=\sigma$, $A(\sigma) = 0$ 
and $B(\sigma) = \sigma$ in equation (31) of L92.
The variable  $\pi$ denotes the hydrostatic pressure, and 
$\pi_s$, the surface hydrostatic pressure.
The system is readily obtained from equations (36) -- (45) of L92 
by removing all horizontal dependencies.
The surface hydrostatic-pressure does not evolve in time 
(see equation (45) in L92).

The equations are linearised around a resting atmospheric-state $\cxba$
and a resting reference-state $\cxst$, both satisfying the conditions of
section \ref{sec_fram}.
The temperatures $\tba$ and $T^*$ are taken uniform, 
and we still define the "non-linearity" factor by: $\alpha = (\tba - T^*)/T^*$.
The pressure values $\pba$ and $p^*$ are assumed to be
equal to a common value $\pi_0$ at the origin ($\sigma=1$). 
Since $\cxba$ and $\cxst$ are hydrostatically balanced, 
$\pba=\pib = \sigma \pi_0$ and $p^* = \pi^* = \sigma \pi_0 $ at any level. 
The thermodynamics equation decouples and the linear system around $\cxba$
for the vertical velocity $w$ , and the pressure deviation 
$p'=p-\pba$ writes in standard notations:

\vspace{-0.4in} 
\begin{eqnarray}
\frac{\dr w}{\dr t} & = & \frac{g}{\pi_0} \frac{\dr p'}{\dr \sigma} 
\label{eq_exaa}\\
\frac{\dr p'}{\dr t} & = & \frac{C_p}{C_v} \frac{g \pi_0}{R \tba} \sigma^2 \frac{\dr w}{\dr \sigma} 
\label{eq_exab}
\end{eqnarray} 
\vspace{-0.41in}

\noindent The same derivation holds for $\clst$ and leads to an operator
formally identical to the RHS of (\ref{eq_exaa})-(\ref{eq_exab}), still
acting on $(w, p')$, but with $\tba$ replaced by $T^*$.
The solution of (\ref{eq_eigen}), implies that 
if $\tba<0$, $\lamba \in \; i \Real \; \Longrightarrow \; w = \widehat{w}\; \sigma^{r}$
with $r < -1$ or $r > 0$.  For the mode with $r<-1$, the 
energy-density is not bounded when $\sigma \rightarrow 0$.
If $\tba\geq 0$, the structure of the normal modes of 
(\ref{eq_exaa})-(\ref{eq_exab}) is given by:

\vspace{-0.4in} 
\begin{eqnarray}
w(\sigma) & = & \widehat{w} \; \sigma^{(i \nu - 1/2)}  = \widehat{w} \; f_1(\sigma) \\
p'(\sigma) & = & \widehat{p'} \; \sigma^{(i \nu + 1/2)}  = \widehat{p'} \; f_2(\sigma) 
\end{eqnarray} 
\vspace{-0.41in}

\noindent where $\nu$ is a real number, and they have a bounded energy-density.
The condition [C2'] therefore requires $\tba\geq 0$ (i.e. $\alpha \geq  -1$).
Finally, [C4] is trivially checked to be satisfied.

\noindent For a 3-TL SI scheme, (\ref{eq_detm}) writes:

\vspace{-0.4in} 
\begin{equation}
\lp \frac{\lambda^2 -1}{2 \Delta t} \rp^2 
= - \frac{(\nu^2 + 1/4) c^{*2}}{ 4 H^{*2}} 
\lp \lambda^2 + 1 \rp 
\lp \lambda^2 + 1 - \frac{2 \alpha \lambda}{1 + \alpha}\rp,
\label{eq_unD_TTL}
\end{equation} 
\vspace{-0.41in}

\noindent where $c^*= \sqrt{(C_p/C_v) R T^*}$ and $H^*=R T^* /g$.
Comparison of (\ref{eq_unD_TTL}) and (\ref{eq_SW_TTL}) shows that
the stability of the 1D vertical system for $\alpha$ is the same as
the one of the previous shallow-water system for $\alpha' = - \alpha / (1 + \alpha)$.
Hence the criteria for unconditional stability 
directly follows from those of the previous case, by similarity 
arguments.
The criterion for unconditional stability of the 3-TL SI scheme 
is $\alpha \geq  (-1/2)$, i.e. $\tba \geq  (1/2) T^*$, and this criterion 
remains unchanged when increasing $N_{\rm iter}$.

For a 2-TL SI non-extrapolating ($\mu=1$) scheme, (\ref{eq_detm}) 
becomes:

\vspace{-0.4in} 
\begin{equation}
\lp \frac{\lambda -1}{\Delta t} \rp^2 
= - \frac{(\nu^2 + 1/4) c^{*2}}{4 H^{*2}} 
\lp \lambda + 1 \rp 
\lp \lambda  + 1 - \frac{2 \alpha}{1 + \alpha} \rp,
\label{eq_unD_DTL}
\end{equation} 
\vspace{-0.41in}

\noindent and the criterion for unconditional stability 
($\alpha\geq 0$, i.e. $\tba \geq  T^*$), which is more constraining 
than for the 3-TL SI scheme. For iterated 2-TL schemes, the 
criterion for  unconditional stability is $\alpha \geq  (-1/2)$
for even values of $N_{\rm iter}$, and  $\alpha \geq  0$ for odd values.

\subsection{1D vertical acoustic system in height-based coordinate}

In this example we show that the stability properties of the
1D vertical system may depend on the coordinate.
The framework is taken as in the previous example except that the
vertical coordinate is the height $z$. 
The linearised system $\clba$ (cf. e.g. Caya and laprise, 1999) 
writes:

\vspace{-0.4in} 
\begin{eqnarray}
\frac{\dr w}{\dr t} & = & - R \tba \frac{\dr q'}{\dr z} + g \frac{T'}{\tba} \\
\frac{\dr T'}{\dr t} & = & - \frac{R \tba}{C_v} \frac{\dr}{\dr z} w \\
\frac{\dr q'}{\dr t} & = & \lp \frac{g}{R \tba}  - \frac{C_p}{C_v} \frac{\dr }{\dr z} \rp  w, 
\end{eqnarray} 
\vspace{-0.41in}

\noindent where $q'=q - \overline{q}$, $q = {\rm ln}(p/p_0)$, 
$\overline{q} = -gz / R \tba$, $p_0$ is a reference pressure,
and $p$ is the true pressure.
The normal modes of $\clba$ have the following form, for 
$\psi = (w, T', q')$:

\vspace{-0.4in} 
\begin{equation}
\psi = \widehat{\psi} \; \exp \lc \lp  i \nu + \frac{1}{2 \hba} \rp z \rc
\end{equation} 
\vspace{-0.41in}

\noindent where $\hba = (R \tba / g)$. The reference system $\clst$ is defined
in a similar way replacing $\tba$ by $T^*$ and $\overline{q}$ by $q^*= -gz/R T^*$.
It should be noted that the structure of the normal modes of $\clst$ 
is not the same as for $\clba$ since the characteristic height for
$\clst$ is $H^* = (R T^* / g)$

For a 2-TL SI non-extrapolating ($\mu=1$) scheme, (\ref{eq_detm}) 
becomes:

\vspace{-0.4in} 
\begin{equation}
\lp \frac{\lambda -1}{\Delta t} \rp^2 
= \frac{c^{*2}}{4} \lp i \nu + \frac{1}{2 \hba} \rp  \lp \lambda + 1 + 2 \alpha \rp 
\lc  i \nu \lp \lambda + 1 \rp - \frac{1}{H^*} \lp \frac{1+ 2 \alpha}{1+\alpha}\rp
\lp \lambda + 1 - \frac{4 \alpha}{1 + 2 \alpha }\rp \rc
\end{equation} 
\vspace{-0.41in}

\noindent where $c^*=\sqrt{(C_p/C_v) R T^*}$. In the height-coordinate framework, 
the asymptotic
growth-rate depends on the structure since $\nu$ appears in one of the factors
which become dominant at large time-steps. The interpretation of the
results is thus slightly complicated in comparison to the case of a
mass-based coordinate. 
For the most external structure ($\nu=0$), the asymptotic growth-rate is given
by the roots of:

\vspace{-0.4in} 
\begin{equation}  
\lp \lambda + 1 + 2 \alpha \rp  \lp \lambda + 1 - \frac{4 \alpha}{1+ 2 \alpha} \rp = 0.
\end{equation} 
\vspace{-0.41in}

\begin{figure}[p]
\epsfxsize=\figwidth
\centerline{\epsfbox{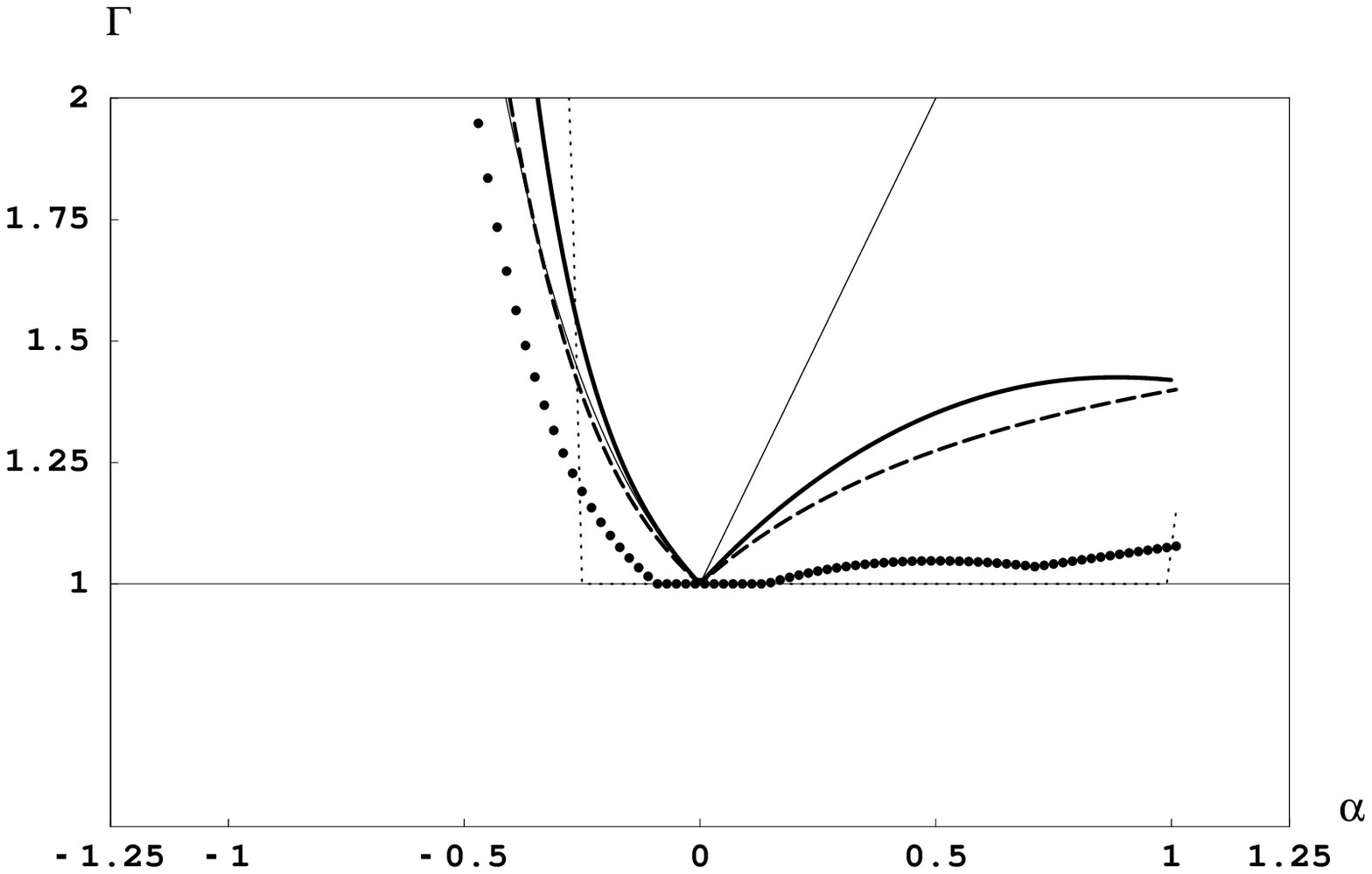}}
\caption{\label{fig_1dz} Asymptotic growth-rates $\Gamma$
 for 1D vertical system in $z$ coordinate
 as a function of the nonlinearity parameter $\alpha$.
  thin line: long mode ($\nu=0.0001 \; {\rm m}^{-1}$) with 2-TL SI scheme;
  thick line:  long mode with 2-TL ICI scheme $N_{\rm iter} = 2$;
  dotted line: external mode ($\nu=0$)with 3-TL SI scheme;
  dashed line: long mode with 3-TL SI scheme.
  Circles: practical growth-rate of the 3-TL SI scheme for the long 
  mode with $\Delta t = 30$~s and $\epsilon=0.1$.
  }
\end{figure}

\noindent This polynomial consists in a combination of 
two factors similar to those obtained in the two previous examples
(through a formal replacement of $2 \alpha$ by 
$\alpha$ for the second factor).
As a consequence, the unconditional stability domains can 
be readily deduced from these previous examples: the external 
structure $\nu=0$ is unstable for any value $\alpha \neq 0$ when 
$\Delta t \longrightarrow \infty$. The instability is thus
much more severe than in the case of a mass-based vertical 
coordinate, for which $\alpha \geq  0$ was sufficient to ensure
unconditional stability. 
Moreover, slightly shorter structures with vertical 
wavelengths of the order of $(1/\hba)$ are found to be more 
unstable than the external one, and for these structures, the
unconditional stability criterion ($\alpha=0$) remains unchanged 
when $N_{\rm iter}$ is increased.
Fig. \ref{fig_1dz} shows the asymptotic growth-rates for
the 2-TL SI $(N_{\rm iter}=1)$ scheme and the the 2-TL
ICI scheme with $N_{\rm iter}=2$ for $\nu=0.0001\; {\rm m}^{-1}$,
a structure for which the instability is close to its maximum.
The severe instability of the 2-TL SI scheme is only alleviated 
but not suppressed by choosing $N_{\rm iter}=2$.

For the 3-TL SI scheme, the external structure $\nu=0$ is unconditionally 
stable for $-0.25 \leq  \alpha \leq  1$, but slightly shorter 
structures as above are found unstable at large time-steps 
as soon as $\alpha  \neq 0$ (very short modes are stable however).
Fig. \ref{fig_1dz} depicts the asymptotic growth-rates for two
structures: the external structure $\nu=0$, and a long
structure $\nu=0.0001\; {\rm m}^{-1}$.
The growth-rate of the long structure
for a moderate time-step $\Delta t = 30$~s with a time-decentering 
$\epsilon=0.1$ (as in Caya and Laprise, 1999)
is also depicted: the practical instability becomes
small in these conditions, and the 3-TL scheme cannot be 
positively rejected, especially considering
the fact that dissipative processes could act in a 
way to stabilize the scheme.
The practical impact of this predicted weak instability for NWP 
applications could be easily evaluated with a $z$-coordinate model, using an 
experimental set-up similar to the one used here, and then 
progressively extending the set-up to approach real-case 
experimental conditions.

\subsection{Comments}

In the three simple examples examined above, the criterion for unconditional 
stability is seen to be more constraining in the 2-TL non-extrapolating SI
scheme than in the 3-TL SI scheme. 
The 2-TL extrapolating SI scheme is found to
have similar domains of unconditional stability
than its non-extrapolating counterpart (not shown).
For mass-based coordinates, if both vertically propagating 
acoustic waves and external gravity waves 
are simultaneously allowed by a given equation system,
the above analyses suggest that 2-TL SI schemes are so 
constraining that there is no domain for unconditional stability. 
For height-based coordinates, the long vertically propagating
acoustic waves are always unstable in the 2-TL SI scheme.

This leads to suspect that, in opposition to 3-TL SI schemes, 
classical 2-TL SI schemes are not suitable for the EE system
with any  vertical coordinate.
The HPE system with 2-TL SI scheme did not suffer from this problem 
since vertically propagating waves are not allowed in HPE (i.e. the 1D 
column atmosphere is stationary).
The intrinsic instability of the 2-TL SI scheme for EE system is confirmed
in section \ref{sec_eepc} for mass-based coordinates.

When a second-order time-filter with parameter $\kappa$ 
is applied to the system examined in the first example
for a 2-TL SI non-extrapolating scheme, (\ref{eq_SW_DTL}) 
becomes:

\vspace{-0.4in} 
\begin{equation}
\lp \frac{\lambda -1}{\Delta t} \rp^2 
= - \frac{k^2 c^{*2}}{4} 
\lc \lp \lambda + 1 \rp + \kappa  \lp \lambda -2 + \frac{1}{\lambda} \rp \rc
\lc \lp \lambda  + 1 + 2\alpha \rp + \kappa  \lp \lambda -2 + \frac{1}{\lambda} \rp \rc,
\end{equation} 
\vspace{-0.41in}

\noindent and the criterion for unconditional stability 
becomes  $-1 \leq  \alpha \leq  2 \kappa$.
For the second example, the similarity argument
shows that the criterion for unconditional stability 
becomes: $\alpha\geq  - 2 \kappa / (1 + 2 \kappa)$.
The domains of stability 
of 3-TL and 2-TL time-filtered ICI schemes for the two first
examples are summarized in Table 1.

The application of a time-filter thus allows to alleviate
the stability constraints for 2-TL SI schemes, and a non-vanishing
domain for unconditional stability is recovered.
The width of the unconditional stability domain
increases with $\kappa$. This is found to hold for the 1D vertical system
in height-based coordinates as well, which is 
consistent with the results of Semazzi et al. (1995)
and Qian et al (1998): they succeeded to solve numerically the EE system
at low-resolution with a 2-TL SI scheme, however, 
the use of a large value $\kappa = 0.5$ was 
required to stabilize the model. As a consequence the
forecasts suffered from a dramatic loss of energy 
with increasing forecast-range, and 
ceased to be of meteorological interest after 2-3 days.
Moreover, at high resolutions (and consequently steep orography)
the use of a time-filter $\kappa$ is found experimentally 
to be an insufficient solution for eliminating the intrinsic 
instability of the scheme (not shown).

If a high level of accuracy is desired for the EE system
with a 2-TL classical SI time-discretization and high resolution, 
a more robust scheme (e.g. with a larger value of 
$N_{\rm iter}$) must be used. 
The above 
analyses show that the unconditional stability 
domain is dramatically reduced for odd values 
of $N_{\rm iter}$, hence ICI schemes with
even values of $N_{\rm iter}$ are preferable
for solving the EE system with a 2-TL scheme.

The 1D vertical system in mass-based coordinates
has been found to be more stable than its counterpart in
height-based coordinates in a general way.
For mass-coordinates, the 3-TL SI and the 2-TL ICI schemes
with even values of $N_{\rm iter}$ have an extended domain
of unconditional stability, whilst for height-based coordinates,
they are unstable as soon as $\alpha \neq 0$.
For 3-TL SI schemes, the necessity to have recourse to a first-order 
time-decentering $\epsilon > 0$ to overcome this instability
is a significant drawback since it results in a spurious
damping of transient phenomena, similarly to $\kappa$
but in an even less selective way, as mentioned above.
We think these differences give a substantial theoretical 
advantage to mass-based coordinates for solving the EE system
with classical SI or ICI schemes.

\section{Analysis of the EE system for isothermal atmospheres in mass-coordinate}
\label{sec_eepc}

The analysis of the isothermal HPE system for ICI schemes does not
substantially modify the general conclusions drawn for the
shallow-water case (not shown), hence the case of the EE
system is directly examined.

In this section, the EE system is cast in the pure 
unstretched terrain-following coordinate  $\sigma$ which can be 
 classically derived from the hydrostatic-pressure
coordinate $\pi$ of L92, through 
$\sigma = (\pi/\pi_s) \in [0,1]$, where $\pi_s$ is the hydrostatic
surface-pressure.  
The nonhydrostatic prognostic variables are
the non-dimensionalised nonhydrostatic 
pressure departure $\pca= (p - \pi) / \pi$
(where $p$ is the true pressure), 
and the vertical divergence $\nd$ which writes
in $\sigma$ coordinate:

\vspace{-0.4in} 
\begin{equation}
\nd   =   - \frac{g}{R T} (1 + \pca) \sigma \frac{\dr w}{\dr \eta}
\label{eq_def_d}
\end{equation} 
\vspace{-0.41in}

The adiabatic system writes:

\vspace{-0.4in} 
\begin{eqnarray}
\frac{d {\bf V}}{dt} & = & - RT \nabla q 
                       -\frac{RT}{(1 + \pca)}  \nabla \pca 
                       - \lp 1 + \pca + \sigma \frac{\dr \pca}{\dr \sigma} \rp \nabla \phi 
  \label{eq_pronos_mod_V}\\
\frac{d \nd}{dt}  & = & 
       - \frac{g^2 (1+\pca)}{RT} \lp \sigma \frac{\dr}{\dr \sigma} \rp 
               \lp 1+ \sigma \frac{\dr}{\dr \sigma} \rp \pca \nonumber \\
      & + &  \nd (\nabla {\bf V} - D_3)           
             +  \frac{g (1+\pca)}{RT} \lc \nabla w \lp \sigma \frac{\dr {\bf V}}{\dr \sigma} \rp \rc 
\label{eq_pronos_mod_d} \\
\frac{dT}{dt} & = & - \frac{RT}{C_v} D_3 
       \label{eq_pronos_mod_T}  \\
\frac{d \pca}{dt} & = &
        - \lp 1 + \pca \rp \lp \frac{C_p}{C_v}   D_3 +  \frac{\dot{\pi}}{\pi} \rp
       \label{eq_pronos_mod_P}\\
\frac{\dr q}{\dr t} & = & - \int_0^1{ \lp \nabla {\bf V} + {\bf V} \nabla q \rp  } d\sigma'
       \label{eq_pronos_mod_q}
\end{eqnarray} 
\vspace{-0.41in}

\noindent where:

\vspace{-0.4in} 
\begin{eqnarray}
D_3 & = & \nabla {\bf V} + \nd 
       + \frac{(1 + \pca)}{RT} \nabla \phi . \lp \sigma \frac{\dr {\bf V}}{\dr \sigma} \rp \\
\phi & = & R \int_{\sigma}^1{ \lp \frac{T}{1 + \pca} \rp  } \frac{d\sigma'}{\sigma'}  \\
\frac{\dot{\pi}}{\pi} & = & {\bf V} \nabla q 
   - \frac{1}{\sigma} \int_0^{\sigma}{ \lp \nabla {\bf V} + {\bf V} \nabla q  \rp  } d\sigma',
\end{eqnarray} 
\vspace{-0.41in}

\noindent {\bf V} is the horizontal wind, and $\nabla$ is the horizontal derivative operator.
The domain is
restricted to a vertical plane along $(x, \sigma)$ directions for clarity.
The system is linearized around a resting isothermal and 
hydrostatically-balanced state $\cxba$: 

\vspace{-0.4in} 
\begin{eqnarray}
\frac{\dr D}{\dr t} & = &  - R {\cal G} \nabla^2 T 
       +  R \tba ({\cal G} - {\cal I} ) \nabla^2 \pca 
       - R \tba \nabla^2 q
\label{eq_linD1} \\
\frac{\dr \nd }{\dr t} & = & 
         - \frac{g^2}{R \tba} \lp 1 + \sigma \frac{\dr}{\dr \sigma} \rp 
            \lp \sigma \frac{\dr}{\dr \sigma} \rp \pca \\
\frac{\dr  T}{\dr t}  & = & - \frac{R \tba}{C_v} ( D  + \nd) 
 \label{eq_linT1}  \\
\frac{\dr  \pca }{\dr t} & = & {\cal S} D 
       - \frac{C_p}{C_v} ( D  + \nd ) 
\label{eq_linP1}  \\
\frac{\dr  q }{\dr t} & = & - {\cal N} D, 
\label{eq_linq1}  
\end{eqnarray} 
\vspace{-0.41in}

\noindent where the vertical integral operators ${\cal G}$, ${\cal S}$ 
and  ${\cal N}$ are defined by:

\vspace{-0.4in} 
\begin{eqnarray}
{\cal G} X & = & \int_\sigma^1{(X/\sigma')} d\sigma' \\
{\cal S} X & = & (1/\sigma) \int_0^\sigma{X} d\sigma' \\
{\cal N} X & = & \int_0^1{X} d\sigma' 
\end{eqnarray} 
\vspace{-0.41in}

\noindent The $\clst$ operator is similar to the RHS of this system, 
simply replacing $\tba$ by $T^*$.

\subsection{Verification of conditions [C1] -- [C4]}

The linear operator $l_1= \sigma ( \dr / \dr \sigma)$ is applied
to (\ref{eq_linD1}), and $l_4 = [ {\cal I}+ \sigma ( \dr / \dr \sigma)]$ 
to (\ref{eq_linP1}). The $q$ equation (\ref{eq_linq1}) decouples and
we obtain a linear unbounded system, in which (\ref{eq_linD1}) and
(\ref{eq_linP1}) are replaced by:

\vspace{-0.4in} 
\begin{eqnarray}
\lp \sigma \frac{\dr }{ \dr \sigma} \rp \frac{\dr D}{\dr t} & = &  R \nabla^2 T 
       -  R \tba \lp  \sigma \frac{\dr }{ \dr \sigma} + {\cal I} \rp \nabla^2 \pca 
\label{eq_linD2} \\
\lp  \sigma \frac{\dr }{ \dr \sigma} + {\cal I} \rp \frac{\dr  \pca }{\dr t} & = & D 
       - \frac{C_p}{C_v} \lp  \sigma \frac{\dr }{ \dr \sigma} + {\cal I} \rp \lp D + \nd \rp
\label{eq_linP2}  
\end{eqnarray} 
\vspace{-0.41in}

\noindent Hence we have $P=4$, $\cx = (D, \nd, T, \pca)$. 
Using of the same operators ($l_1$, $l_4$), the
reference operator is also made free of any reference to the 
upper and lower boundary conditions, which
shows that the condition [C1] is satisfied.
Solution of (\ref{eq_eigen}) shows that [C2] requires $\tba \geq  0$ (i.e. $\alpha \geq  -1$).
The normal modes of the system are then:

\vspace{-0.4in} 
\begin{equation}
\psi (x,\sigma)  =  \widehat{\psi} \; \exp(ikx)\, \sigma^{(i \nu - 1/2)}  
\end{equation} 
\vspace{-0.41in}

\noindent where $(k,\nu) \in \Real$ and $\psi$ represents $D$, $\nd$, $T$ or $\pca$. 
In this particular case, the $f_1, \ldots, f_4$ functions are all identical.
The verification of [C3], [C4] proceeds easily, as in previous sections.

\subsection{Results}

\begin{figure}[p]
\epsfxsize=\figwidth
\centerline{\epsfbox{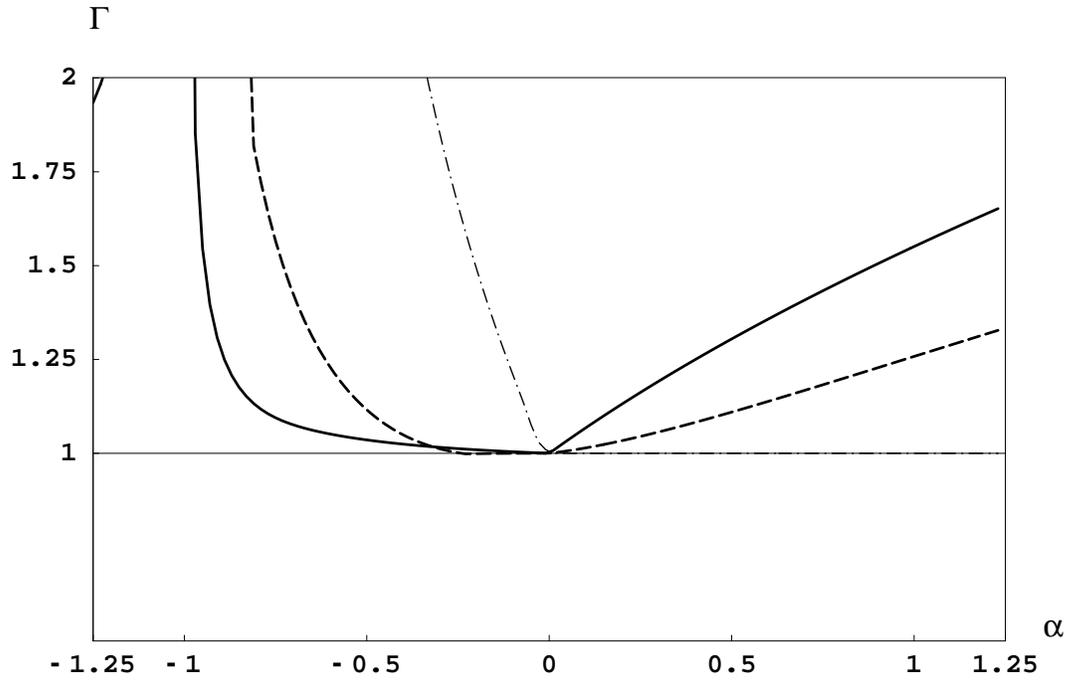}}
\caption{\label{fig_eemodes} Growth-rate  $\Gamma$
 with $\Delta t = 20~s$ for the EE system with a 2-TL SI scheme
 as a function of the nonlinearity parameter $\alpha$.
  solid line: external mode (i);
  dashed line: slantwise mode (ii);
  dot-dashed line: internal mode (iii).
  The left-part of the solid line represents an acoustic external mode.
  }
\end{figure}

As a first illustration of the results, the growth-rates of 
2-TL non-extrapolating ICI schemes are shown in Fig. \ref{fig_eemodes} as function of 
$\alpha = (\tba - T^*)/T^*$ with a moderate time-step $\Delta t = 20~s$, for three
particular mode structures: 

(i) an external mode ($k=0.0005~{\rm m}^{-1}$, $\nu=0$)

(ii) a vertical very internal mode ($k=0$, $\nu=100$)

(ii) an intermediate slantwise mode ($k=0.0005~{\rm m}^{-1}$, $\nu=3$)

\noindent The suspicions raised in the simple 1D examples are 
confirmed: the internal vertically-propagating mode is unstable
for $\alpha < 0$ while the external gravity mode is unstable
for $\alpha>0$. Moreover, intermediate, slantwise-propagating
modes are unstable in the whole domain, and the acoustic external
mode (Lamb-wave) appears to be  unstable for $\alpha<0$ as well.
The domain of stability vanishes, which confirms that 2-TL SI 
scheme is not relevant for solving the EE system. The effect
of introducing a time-filter to save the situation is discussed
below.

\begin{figure}[p]
\epsfxsize=\figwidth
\centerline{\epsfbox{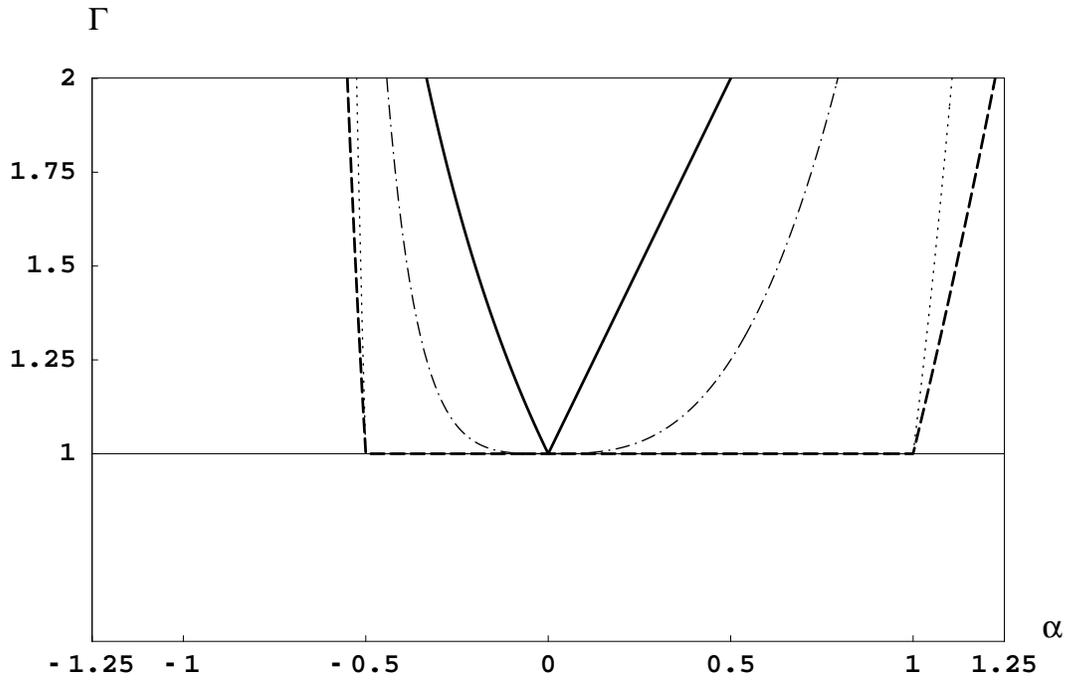}}
\caption{\label{fig_eeici} 
Asymptotic growth-rates $\Gamma$
 for EE system with 2-TL ICI scheme as a function of the 
 nonlinearity parameter $\alpha$.
  solid line: $N_{\rm iter} = 1$;
  dashed line: $N_{\rm iter} = 2$;
  dot-dashed line: $N_{\rm iter} = 3$;
  dotted line: $N_{\rm iter} = 4$.
  }
\end{figure}

The asymptotic growth-rates resulting from the EE system 
for $\Delta t = \infty$ are now examined. Similarly to most previous
cases, they are independent of the geometry ($k$, $\nu$) of the mode.
Fig. \ref{fig_eeici} shows the asymptotic growth-rates 
as a function of $\alpha$ for 2-TL non-extrapolating 
ICI schemes with $N_{\rm iter} = (1, 2, 3, 4)$.
As stated above, the SI scheme ($N_{\rm iter} = 1$)
is unstable for any value of $\alpha$.
For even values of $N_{\rm iter}$, the scheme has an "optimal"
domain of unconditional stability $-1/2 \leq  \alpha \leq  1$, while
for odd values, the scheme is unstable for all values of $\alpha$.

For 3-TL ICI schemes the domain of unconditional stability is 
$-1/2 \leq  \alpha \leq  1$ independently of the values of $N_{\rm iter}$ 
and  $\kappa$. 
The curves (not shown) are similar to those obtained for even values of  
$N_{\rm iter}$ for 2-TL ICI schemes.

\begin{figure}[p]
\epsfxsize=\figwidth
\centerline{\epsfbox{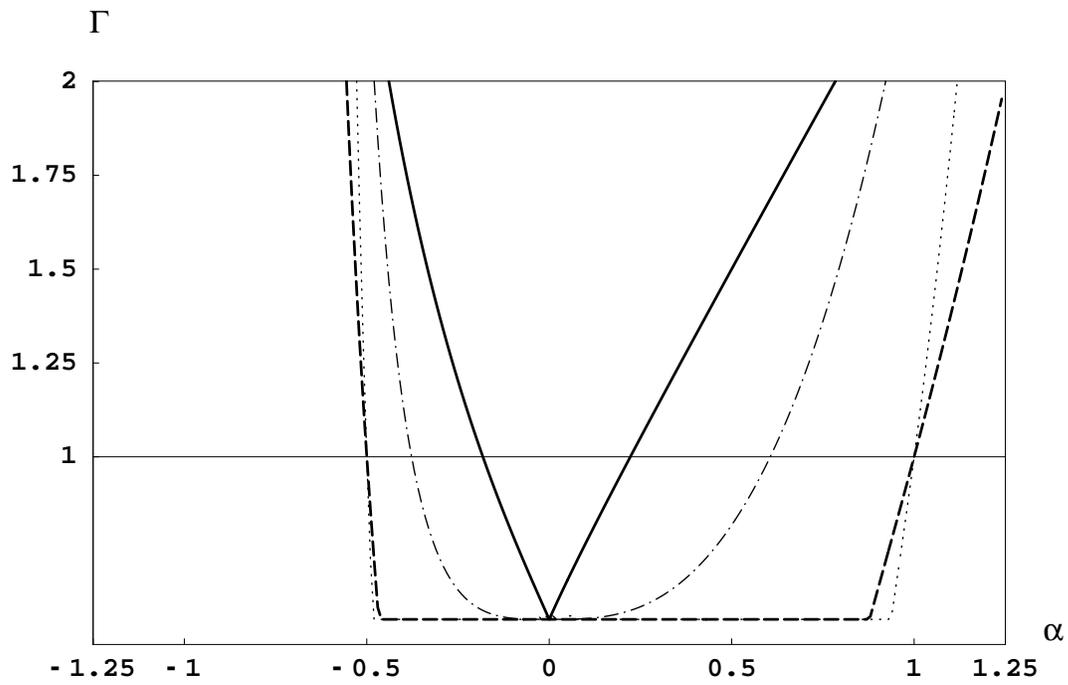}}
\caption{\label{fig_eeicik} 
Same as Fig 3, but with a time-filter $\kappa=0.1$.
  solid line: $N_{\rm iter} = 1$;
  dashed line: $N_{\rm iter} = 2$;
  dot-dashed line: $N_{\rm iter} = 3$;
  dotted line: $N_{\rm iter} = 4$.
  }
\end{figure}

The impact of applying a time-filter $\kappa=0.1$ to 
2-TL ICI schemes is depicted on Fig \ref{fig_eeicik}
for the  second variant (the first variant behave qualitatively  
in the same way).
The global impact is to "lower" the curves
of the asymptotic growth-rates, and consequently, to increase
the width of the unconditional-stability domain. However, large
values of $\kappa$ (e.g. $\kappa \approx 0.5$) are required in order 
to obtain a wide stability domain, especially for the 2-TL SI scheme,
and this strategy is known to be irrelevant for NWP purpose.

Finally, it is worth noting that the results obtained for the EE system
are fully compatible with the conclusions that can be drawn from
the intersection of the domains of unconditional stability in Table 1
for the two simple frameworks examined above in mass-coordinate.  
The ability of these very refined frameworks to capture the essence 
of the behaviour of the time-discretised EE system in the limit of 
long time-steps makes them very useful tools to fully understand the underlying 
causes of its stability or instability.

\section{Conclusion}

A general method for investigating the stability of 
the ICI class of time-discretisations on canonical problems
with various space-continuous equation systems
has been presented. These ICI schemes are based on 
a separation of evolution terms between a simple linear 
operator and "non-linear" residuals.
The method has been validated by confirming 
earlier results, then the application to new frameworks 
(equation systems or 
time-discretisation schemes) allowed to extend these results.
The main conclusions drawn from this study are:

\begin{list}{}{}
\item[(i)] Even on very simple (1D) examples, the stability
properties of time-discretisations for a given equation system
are very dependent on fundamental choices (as e.g. the choice 
of the vertical coordinate). 
Hence, the import of conclusions drawn from a given analysis 
must be carefully limited to the examined framework. 
\vspace{-0.2in}

\item[(ii)] For the EE system,
height-based coordinates have a theoretical disavantage
compared to mass-based coordinates since they exhibit an 
intrinsic instability for (long) vertically propagating waves.
\vspace{-0.2in}

\item[(iii)] The 2-TL SI scheme is found not to be appropriate 
for the EE set of equations, whatever coordinate is employed
(using a time-filter results in an unacceptable 
degradation of the solution).
\vspace{-0.2in}

\item[(iv)] For the EE system, the 2-TL scheme with
$N_{\rm iter} = 2$ brings a dramatic increase of the 
stability compared to the 2-TL SI scheme ($N_{\rm iter} = 1$).
This statement holds for even values of $N_{\rm iter}$, 
while odd values leads to a significantly weaker stability.
\vspace{-0.2in}

\item[(v)] As a consequence of the latter point, the 2-TL
ICI scheme with $N_{\rm iter} = 2$ seems worth to be
considered for the EE system.

\end{list}

\vspace{-0.2in}

However, as mentioned in SHB78, the stability
inferred from this type of analyses is overestimated, 
and flows in which the non-linearity comes from other
sources than the discrepancy between the atmospheric and reference
temperature profiles could reveal new instabilities in
practice. For instance, in spite of its apparent "optimal" stability
in the simplified context of this paper, the 3-TL SI scheme has proved
to be not stable enough for solving numerically the EE system
in realistic highly non-linear conditions at high resolutions, due to
other terms treated explicitly.
This point clearly demonstrates the limitations of 
this type of academic exercise.

Nevertheless, in spite of its necessary limitations,
this study can serve to distinguish 
schemes which are definitely not relevant for practical use
from the others, and give a first theoretical 
justification for those which are worth considering. 
In agreement with Cote et al (1998) and Cullen (2000)
we think that ICI schemes with $N_{\rm iter} \geq 2$ are 
among the most approriate alternatives for integrating
the EE system in highly non-linear conditions at fine-scale,
including from the point of view of efficiency.

\newpage
\section*{References}

\begin{description}

\item Bubnov\'a, R., G. Hello, P. B\'enard, and J.F. Geleyn, 1995:
      Integration of the fully elastic equations cast in the hydrostatic
      pressure terrain-following coordinate in the framework of the
      ARPEGE/Aladin NWP system.
      {\em Mon. Wea. Rev.}, {\bf 123}, 515-535.
      
\item Caya, D., and R. Laprise, 1999:
      A semi-implicit semi-Lagrangian regional climate model: the Canadian RCM.
      {\em Mon. Wea. Rev.}, {\bf 127}, 341-362.
      
\item C\^{o}t\'{e}, J., M. B\'eland, and A. Staniforth, 1983:
      Stability of vertical discretization schemes for semi-implicit
      primitive equation models: theory and application.
      {\em Mon. Wea. Rev.}, {\bf 111}, 1189-1207.
      
\item C\^{o}t\'{e}, J., S. Gravel, A. M\'ethot, A. Patoine, M. Roch, and 
      A. Staniforth, 1998:
      The Operational CMC-MRB Global Environmental
      Multiscale (GEM) Model. Part I: Design Considerations and Formulation.
      {\em Mon. Wea. Rev.}, {\bf 126}, 1373-1395.
\item Cullen, M. J. P., 2000:
      Alternative implementations of the semi-Lagrangian semi-implicit 
      schemes in the ECMWF model.
      {\em Q. J. R. Meteorol. Soc.}, {\bf 127}, 2787-2802.
\item Hereil, P., and R. Laprise, 1996:
      Sensitivity of Internal Gravity Waves Solutions to the
      Time Step of a Semi-Implicit Semi-Lagrangian Nonhydrostatic Model.
      {\em Mon. Wea. Rev.}, {\bf 124}, 972-999.
\item Laprise, R., 1992:
      The Euler equations of motion with hydrostatic pressure as an
      independent variable.
      {\em Mon. Wea. Rev.}, {\bf 120}, 197-207.
\item Qian, J.-H., F. H. M. Semazzi, and J. S. Scroggs, 1998:
      A global nonhydrostatic semi-Lagrangian atmospheric model with
      orography.
      {\em Mon. Wea. Rev.}, {\bf 126}, 747-771.
\item Robert, A. J., J. Henderson, and C. Turnbull, 1972: An implicit time
      integration scheme for baroclinic models of the atmosphere . 
      {\em Mon. Wea. Rev.}, {\bf 100}, 329-335.
\item Semazzi, F. H. M., J. H. Qian, and J. S. Scroggs, 1995:
      A global nonhydrostatic semi-Lagrangian atmospheric model without
      orography.
      {\em Mon. Wea. Rev.}, {\bf 123}, 2534-2550.
\item Simmons, A. J., B. Hoskins, and D. Burridge, 1978:
      Stability of the semi-implicit method of time integration.
      {\em Mon. Wea. Rev.}, {\bf 106}, 405-412.
\item Simmons, A. J., C. Temperton, 1997:
      Stability of a two-time-level semi-implicit integration scheme
      for gravity wave motion.
      {\em Mon. Wea. Rev.}, {\bf 125}, 600-615.      
\item Tanguay, M., A. Robert, and R. Laprise, 1990:
      A Semi-Implicit Semi-Larangian Fully Compressible Regional Forecast Model.
      {\em Mon. Wea. Rev.}, {\bf 118}, 1970-1980.
\end{description}

\newpage
\section*{List of Tables}

Table 1: Domains of unconditional stability for time-filtered
schemes for the two first 1D examples.

\newpage

\vspace{3mm}

\begin{tabular}{|c|c|c|}
\hline
           & Shallow-water    &  1D vertical (mass)  \\
\hline
3-TL ICI   & $-1 \leq \alpha \leq 1$  & $-1/2 \leq  \alpha$ \\
\hline
2-TL ICI  ($N_{\rm iter}$ even) & $-1 \leq \alpha \leq 1$  & $-1/2 \leq  \alpha$ \\
\hline
2-TL ICI  ($N_{\rm iter}$ odd) \rule[-5mm]{0mm}{10mm}
   & $-1 \leq \alpha \leq  (2 \kappa)^{1/N_{\rm iter}}$  &
   $\frac{\textstyle -2 \kappa^{(1/N_{\rm iter})}}
   {\textstyle 1+ 2 \kappa^{(1/N_{\rm iter})}} \leq  \alpha$ \\
\hline
\end{tabular}

\vspace{3mm}

Table 1: Domains of unconditional stability for time-filtered
schemes for the two first 1D examples.

\vspace{3mm}

\newpage
\section*{List of Figures}

 Fig. 1:  Asymptotic growth-rates $\Gamma$
 for 1D vertical system in $z$ coordinate
 as a function of the nonlinearity parameter $\alpha$.
  thin line: long mode ($\nu=0.0001 \; {\rm m}^{-1}$) with 2-TL SI scheme;
  thick line:  long mode with 2-TL ICI scheme $N_{\rm iter} = 2$;
  dotted line: external mode ($\nu=0$)with 3-TL SI scheme;
  dashed line: long mode with 3-TL SI scheme.
  Circles: practical growth-rate of 3-TL SI scheme for the long mode 
  with $\Delta t = 30$~s and $\epsilon=0.1$.

\medskip

Fig. 2: Growth-rate $\Gamma$
 with $\Delta t = 20~s$ for the EE system with a 2-TL SI scheme
 as a function of the nonlinearity parameter $\alpha$.
  solid line: external mode (i);
  dashed line: slantwise mode (ii);
  dot-dashed line: internal mode (iii).
  The left-part of the solid line represents an acoustic external mode.
  
\medskip

Fig. 3: Asymptotic growth-rates $\Gamma$
 for EE system with 2-TL ICI scheme as a function of the 
 nonlinearity parameter $\alpha$.
  solid line: $N_{\rm iter} = 1$;
  dashed line: $N_{\rm iter} = 2$;
  dot-dashed line: $N_{\rm iter} = 3$;
  dotted line: $N_{\rm iter} = 4$. 
  
 \medskip

Fig. 4: Same as Fig 3, but with a time-filter $\kappa=0.1$.
  solid line: $N_{\rm iter} = 1$;
  dashed line: $N_{\rm iter} = 2$;
  dot-dashed line: $N_{\rm iter} = 3$;
  dotted line: $N_{\rm iter} = 4$.
  
\newpage

\addtocontents{lof}{\protect\vspace{0.5cm}}

\addtocontents{lof}{\protect\vspace{0.5cm}}

\addtocontents{lof}{\protect\vspace{0.5cm}}

%
\end{document}